\documentclass[aps,prl,twocolumn,10pt,nofootinbib,superscriptaddress,floatfix]{revtex4-2}
\usepackage{latexsym}
\usepackage{amsmath}
\usepackage{amsfonts}

\usepackage{hhline}
\usepackage{graphicx}
\usepackage{amssymb}

\usepackage{rotating}
\usepackage{afterpage}

\usepackage{xr-hyper}

\usepackage{color}
\usepackage[normalem]{ulem}

\makeatletter
\newcommand*{\balancecolsandclearpage}{%
  \close@column@grid
  \twocolumngrid
}
\makeatother

\begin{document}
\title{Physical entanglements mediate coherent motion of the 
active topological glass confined within a spherical cavity}
\author{Iurii Chubak}
\email{iurii.chubak@univie.ac.at}
\thanks{Joint first author}
\affiliation{Faculty of Physics, University of Vienna, Boltzmanngasse 5, A-1090 Vienna, Austria}	
\affiliation{Sorbonne Universit{\'e} CNRS, Physico-Chimie des {\'e}lectrolytes et Nanosyst{\`e}mes Interfaciaux, F-75005 Paris, France}
\author{Stanard Mebwe Pachong}
\email{pachong@mpip-mainz.mpg.de}
\thanks{Joint first author}
\affiliation{Max Planck Institute for Polymer Research, Ackermannweg 10, 55128 Mainz, Germany}
\author{Kurt Kremer}
\email{kremer@mpip-mainz.mpg.de}
\affiliation{Max Planck Institute for Polymer Research, Ackermannweg 10, 55128 Mainz, Germany}
\author{Christos N. Likos}
\email{christos.likos@univie.ac.at}
\affiliation{Faculty of Physics, University of Vienna, Boltzmanngasse 5, A-1090 Vienna, Austria}
\author{Jan Smrek}
\email{jan.smrek@univie.ac.at}
\affiliation{Faculty of Physics, University of Vienna, Boltzmanngasse 5, A-1090 Vienna, Austria}

\begin{abstract}
Motivated by chromosomes enclosed in nucleus and the recently discovered active topological glass, we study a spherically confined melt of long nonconcatenated active polymer rings. Without activity, the rings exhibit the same average large-scale conformational properties as chromatin fiber. Upon activating consecutive monomer segments on the rings, 
the system arrives at a glassy steady state due to activity-enhanced topological constraints. The latter generate coherent motions of the system, however the resulting large-scale structures are inconsistent with the fractal globule model.
We observe microphase separation between active and passive 
segments without systematic trends in the positioning of active 
domains within the confining sphere. We find that tank-treading of active segments along the ring contour enhances
active-passive phase separation in the state of active topological
glass when both diffusional and conformational relaxation of the rings are significantly suppressed.
Finally, although the present model of partly-active rings is not compatible with the large-scale chromatin organization, 
our results suggest that the activity-enhanced entanglements that 
result in facilitated intra- and inter-chromosomal contacts 
might be relevant for chromatin structure at smaller scales.
\end{abstract}

\maketitle

The active topological glass (ATG) is a state of matter composed of polymers with fixed, circular, unknotted topology, that vitrifies upon turning a block of monomers within the polymers active and fluidizes reversibly \cite{Active_topoglass_NatComm20}. Unlike classical glasses, where the transition is driven by temperature or density, the ATG results from physical, tight, threading entanglements, generated and maintained by the activity of polymer segments. The activity acting on the ring segments, modeled here as stronger-than-thermal fluctuations, triggers a directed snake-like motion that overcomes entropically unfavorable states and results in significantly enhanced inter-ring threading \cite{Topo2}. A topological glass is hypothesized to exist also in equilibrium solutions of sufficiently long ring polymers, where rings naturally thread (pierce through each other's opening). However, the conjectured critical ring length is currently beyond the experimental or computational reach \cite{Michieletto_Turner_PNAS16,Michieletto_Nahali_Rosa_PRL17,Turner_EPL13}. Although the ATG exhibits accessible critical ring lengths, a formidable challenge in simulating these systems stems from large system sizes that are necessary to avoid self-threading of significantly elongated partly-active rings due to periodic boundary conditions \cite{Active_topoglass_NatComm20}. To overcome this difficulty, a much smaller system confined to an impenetrable cavity can be simulated.

However, in analogy to classical glasses, where the confinement affects the vitrification mechanism and shifts the glass transition temperature in comparison to the bulk value \cite{Classical_confined_vitrification_JNCS91,Classical_confined_vitrification_JPCM94}, it is necessary to ask the question whether the ATG, the existence of which relies on highly extended configurations that promote inter-molecular entanglement, can exist in such a strong confinement at all.  

Besides the ATG, the confined melt of uncrossable polymer rings with active segments has an interesting biological connection. The \emph{equilibrium} melt of rings exhibits conformational properties consistent with the large-scale, population-averaged properties of chromatin fiber in the interphase nuclei of higher eukaryotes \cite{Rosa_Everaers_PloS08,Rings_and_chromosomes_review,ConfRings2020}. In detail, the territorial segregation of distinct chains, the critical exponents $\nu=1/3$ and $\gamma \simeq 1.1$ governing the scaling of the gyration radius $R(s)\sim s^{\nu}$ and the probability of end-contacts $P(s) \sim s^{-\gamma}$ of a segment of length $s$ respectively, coincide for the two systems and characterize the so-called fractal (crumpled) globule conformations \cite{A.Yu.Grosberg1988}. However, similarly to partly active rings, chromatin is out of equilibrium on smaller scales as well. Various processes, such as transcription or loop extrusion inject energy into the system by the action of respective molecular machines on the chromatin fiber. Fluorescence experiments \cite{Zidovska2013} and the related analytical theory \cite{Bruinsma_chromatin_hydrodynamics_BPJ14} suggest that some active events at small scales render fluctuations with thermal spectrum at an effective temperature about twice higher than the ambient one.  
While \emph{not} aiming at a faithful biological representation of the chromatin, we question whether the ATG is consistent with the fractal globule model, since both of the latter represent some aspects of the chromatin conformations in space and time. 

Motivated by both the question on the existence of the confined ATG and the question on the consistency with the fractal globule model of chromatin, here we explore the static and dynamic properties of long, confined, partly-active, nonconcatenated rings in melt. We find that essentially the same phenomenon of the ATG formation is present in confined systems with a small number of polymer chains. The ability to simulate longer rings than in the bulk, allows us to assess in more detail the conformational
and scaling properties of the chains in the non-equilibrium glassy state.
We discover that intermediate-length ring segments feature conformations consistent with a self-avoiding random walk 
($\nu = 0.588$, $\gamma = 1.75$).
While the territorial structure of the fractal globule is clearly distorted, we observe active-passive microphase-separated domains and large-scale correlated motion arising from the glassy phase due to the activity-induced topological constraints. In contrast to
chromatin models in \cite{Hyeon_Thirumalai_PloS18,Liu_correlated_motion_with_lamina_arXiv20}, where the large-scale dynamical coherence arises from explicit interaction potentials or crosslinks, here we show that the activity-induced entanglement can mediate the correlated motion as well. 
Finally, we observe tank-treading of active segments along the ring contour in the glassy state that acts to enhance
active-passive phase separation when both chain diffusion and its conformational rearrangements are suppressed.

We start from a well equilibrated sample of $M = 46$ rings 
each of length $N$ obtained in \cite{ConfRings2020} 
($N = 200, 400, 800$ and $1600$ that corresponds to chain
entanglement lengths of $Z = N/N_{\rm e} \in [7-57]$), the longest being four times longer than the system in \cite{Active_topoglass_NatComm20}. We impose the activity 
on a consecutive segment of length $N/8$ by coupling
it to a (hot) thermostat with
temperature $T_{\rm h} = 3 T_{\rm c}$, where $T_{\rm c}$ is the temperature
of passive (cold) monomers. We use the well-established polymer model~\cite{KG_model,Halverson_statics2011, Halverson_dynamics2011}, 
described in detail in the Supporting Information (SI). The radius $R$ of the confining sphere, which is modeled as a smooth structureless purely repulsive barrier (SI), is fixed by the total monomer density $\rho = 0.85\sigma^{-3}$ for all systems. Typically, $R$ is about $2.5-2.7$ times larger than the equilibrium radius of gyration of the confined chains (Tab.~S3).

\begin{figure}
\centering
\includegraphics{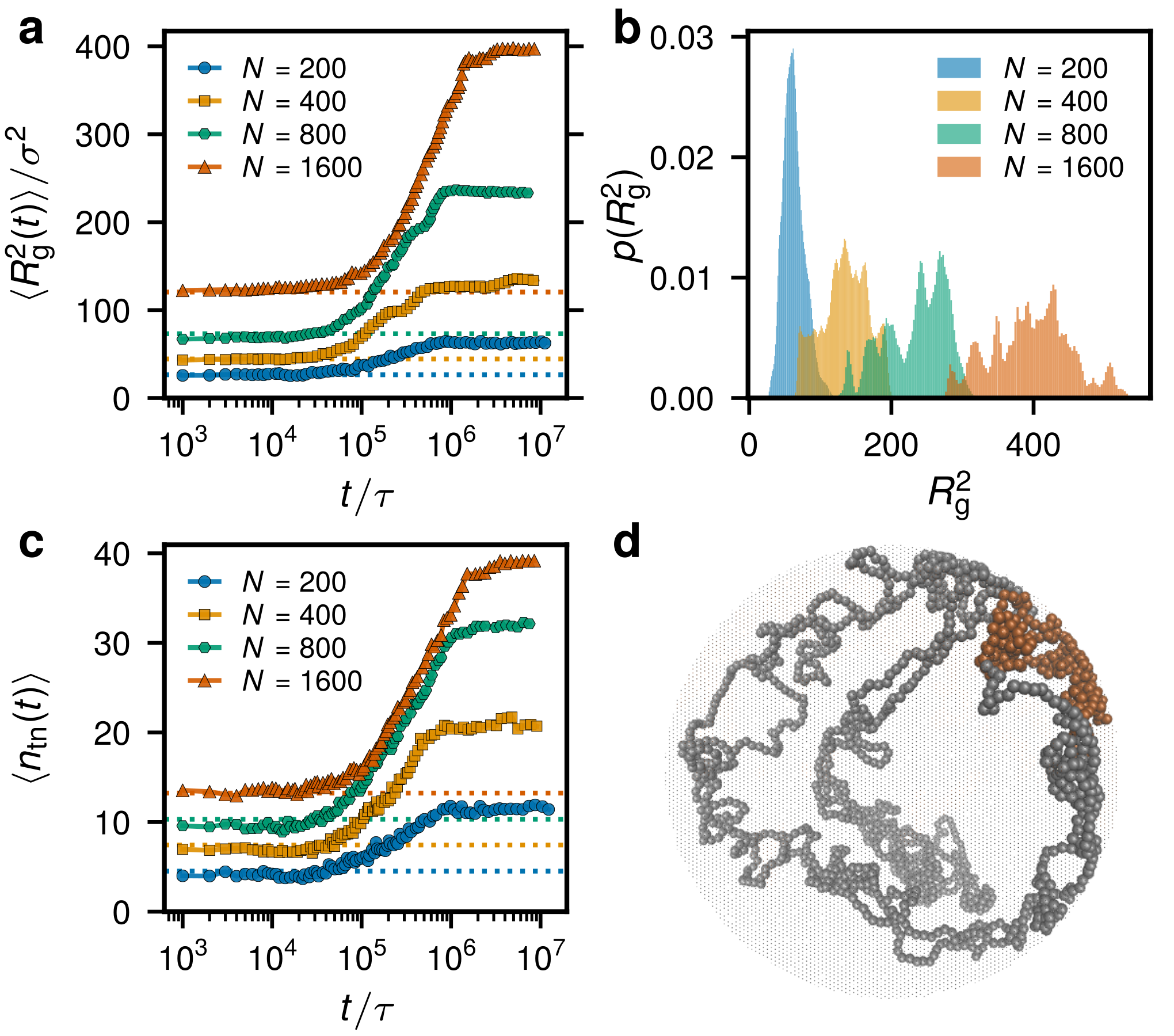}
\caption{\label{fig:statics} {\bf Conformational properties.}
{\bf a}, Evolution of the ring's  $R_{\rm g}^2$ after the activity 
onset at $t = 0$ for systems with different $N$. 
{\bf b}, Distribution of $R_{\rm g}^2$ in the steady state. 
{\bf c}, Evolution of the mean number of threaded neighbors (see Section~S2 in the SI).
{\bf d}, Conformation of a partly-active ring with $N = 1600$
at the end of the simulation run.
In {\bf a} and {\bf c}, the dashed lines of the respective 
color indicate the values in equivalent equilibrium ring melts \cite{ConfRings2020}.
}
\end{figure}

When the activity is switched on, after about $10^{5}\tau$, $\tau$ being the microscopic time of the model (see Section~S1 in the SI), the chains start to expand from their equilibrium sizes until they reach a steady state after $(2-3){\cdot}10^{6}\tau$. The time of the onset of the chain stretching does not significantly depend on $N$ because it is related to local threading constraints. The steady state is characterized by a significantly enhanced mean-square radius of gyration $R_{\textrm{g}}^{2}$ (see snapshot of a chain in Fig.~\ref{fig:statics}{\bf d} and more in Fig.~S1 in the SI, time evolution of $R_{\textrm{g}}^{2}$ in Fig.~\ref{fig:statics}{\bf a} and Tab.~S1 for shape parameter comparison). In comparison to the bulk \cite{Active_topoglass_NatComm20}, the confined rings are significantly less expanded in terms of $R_{\rm g}$ and the ratio of the two biggest eigenvalues of the gyration tensor (Fig.~S2{\bf a} and compare Tab.~S1 to Tab.~S2 of $N=400$). Nevertheless, the conformations are mostly doubly-folded and the change in the shape parameters is due to ``reflections" from the walls. This can be seen in the mean-square internal distance of the longest rings being non-monotonic function of the contour length (Fig.~S2{\bf b}).  In comparison to equilibrium (Tab.~S3), the rings are highly stretched and exhibit self-avoiding walk-like scaling at intermediate distances (Fig.~S2{\bf c}) with a consistent scaling of the contact probability with the exponent $\gamma$ close to $1.75$ (Fig.~S2{\bf d}) \cite{Rings_and_chromosomes_review} and a plateau at largest distances, signifying the loss of correlation due to reflections of rings from the walls. The stretched conformations and the different profile of the contact probability signifies the loss of the original crumpled globule characteristics. The stretching due to the snake-like motion is caused by strong dynamic asymmetry between the active and the passive segments, apparently triggered by non-equilibrium phase separation \cite{Smrek_Kremer_PRL17,Active_topoglass_NatComm20,Topo2}. The dynamics of the mutual ring threading coincides with the stretching dynamics and exhibits markedly enhanced numbers of threaded neighbors  $n_{\textrm{tn}}$ by a single ring in the steady state in comparison to equilibrium (Fig.~\ref{fig:statics}{\bf c}), as we showed by analyzing piercings of rings through other rings' minimal surfaces (SI) \cite{Smrek_Grosberg_minimal_surfaces_ACS16,Smrek_Kremer_Rosa_ACS19,ConfRings2020,Active_topoglass_NatComm20,Tadpoles2020}. Interestingly, the number of threaded neighbors is the same as for the active topological glass in the bulk, despite the different ring shape (compare Tab.~S1 and S2). For the longest rings, each ring practically threads all the other rings in the system. Although this does not hold for the shorter rings, their dynamic behavior is comparable, as detailed below. 

The steady states exhibit a rugged distribution of $R_{\textrm{g}}^{2}$ (Fig.~\ref{fig:statics}{\bf b}), despite averaging over about $10^{7}\tau$, time that is more than one order of magnitude above the equilibrium diffusion times for $N\leq 800$. This shows that the individual chains are not able to change their conformations significantly, being frozen essentially in the same state, and points to a non-ergodic behavior. When averaged over 10 independent runs, a smoother distribution is recovered, as shown for $N=200$ in Fig.~\ref{fig:statics}{\bf b}.

\begin{figure}
\centering
\includegraphics{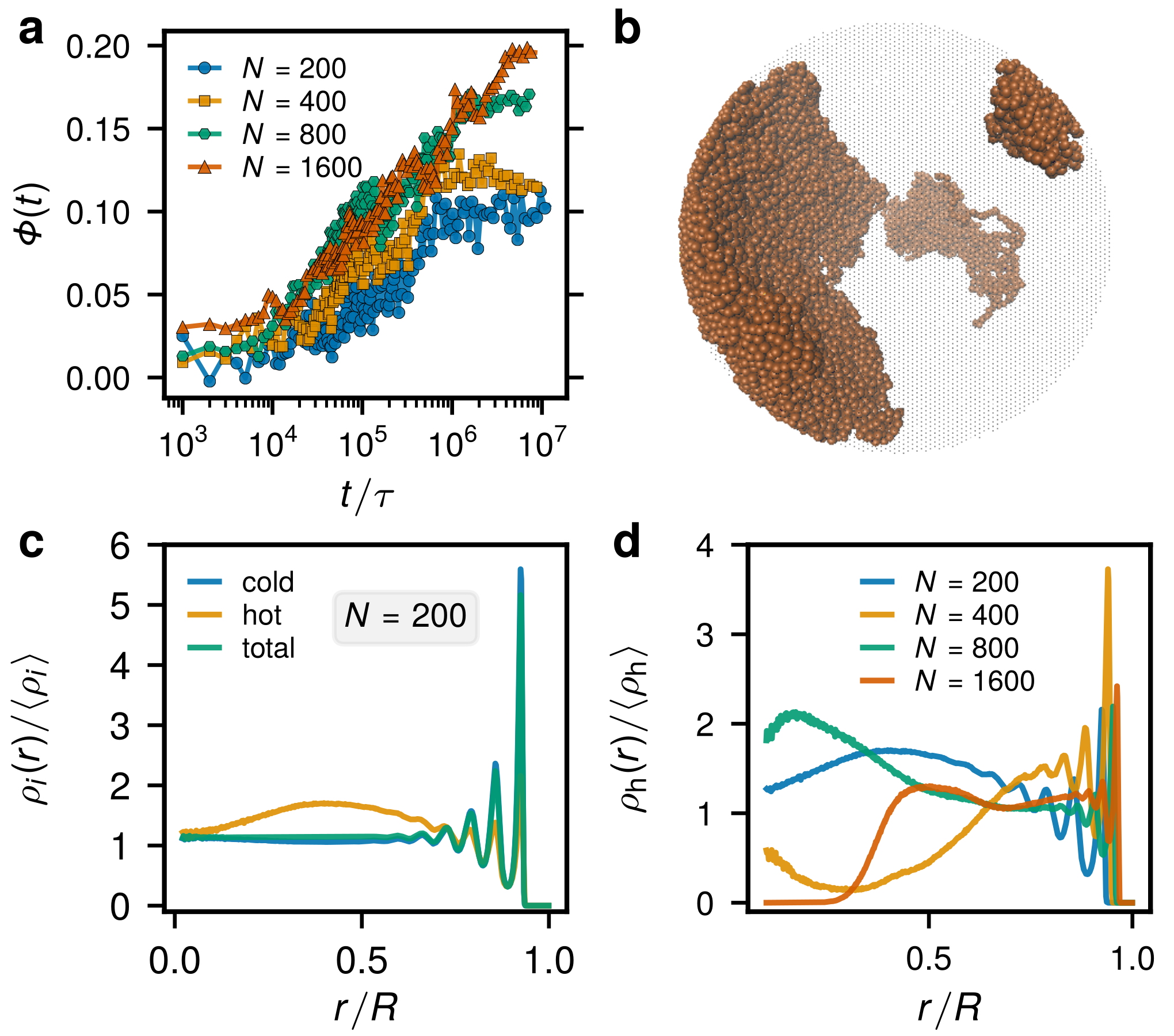}
\caption{\label{fig:phase_sep} {\bf Phase segregation.}
{\bf a}, Time evolution of the order parameter
$\Phi(t)$.
{\bf b}, Phase-segregated regions of hot monomers (cold not shown for clarity) for the
system with $N = 1600$.
{\bf c}, Radial distribution of cold (blue), hot (yellow),
and all (green) monomers within the enclosing sphere for the 
system with $N = 200$ (averaged over 10 independent runs). 
{\bf d}, Radial distributions of hot monomers.}
\end{figure}

The chosen model parameters trigger active-passive (micro)phase separation in all the systems \cite{Topo2,Smrek_Kremer_PRL17,Smrek_Kremer_Entropy18}. We track the degree of phase separation by the order parameter $\Phi(t) = x(t)/x(0) - 1$, where $x(t)$ is the number fraction of inter-chain like-particles in a $r_{\rm{c}} = 2^{1/6}\sigma$ neighborhood of a given monomer at a given time $t$, averaged over monomers (Fig.~\ref{fig:phase_sep}{\bf a}). The initial increase of the order parameter precedes the ring stretching and threading dynamics, supporting the conjecture in \cite{Topo2} that the separation tendency is a precursor of the formation of the glass. The phase separation is dynamic in nature, showing intervals of a single mostly-hot region, but also subsequent dissociation into several hot blobs (Supporting Video 1) reminiscent of the dynamics of activity-driven colloidal crystals \cite{Palacci_living_crystals_Science13}. When the shape properties arrive at a steady state, there are several hot blobs (Fig.~\ref{fig:phase_sep}{\bf b}) and we still observe them occasionally exchanging hot particles. As described below, these are the consequence of a rare tank treading motion of some of the rings, by which the hot segment joins the hot phase without changing the overall shape of the ring and the system as a whole. The radial density distribution of the hot monomers averaged over 10 different runs for $N = 200$ displays confinement induced layering at the wall as in equilibrium \cite{ConfRings2020}, and displays another broad maximum around $R/2$ (Fig.~\ref{fig:phase_sep}{\bf c}). However, the analysis of single runs for $N = 200$ and for other $N$ shows that the positioning of hot monomers is history-dependent, arrested by the topological constraints, and allows for both, internal or peripheral locations (Fig.~\ref{fig:phase_sep}{\bf d}) in contrast to preference for central locations of active monomers in a different polymer model  in \cite{Ganai_NAR14}.

\begin{figure}
\centering
\includegraphics{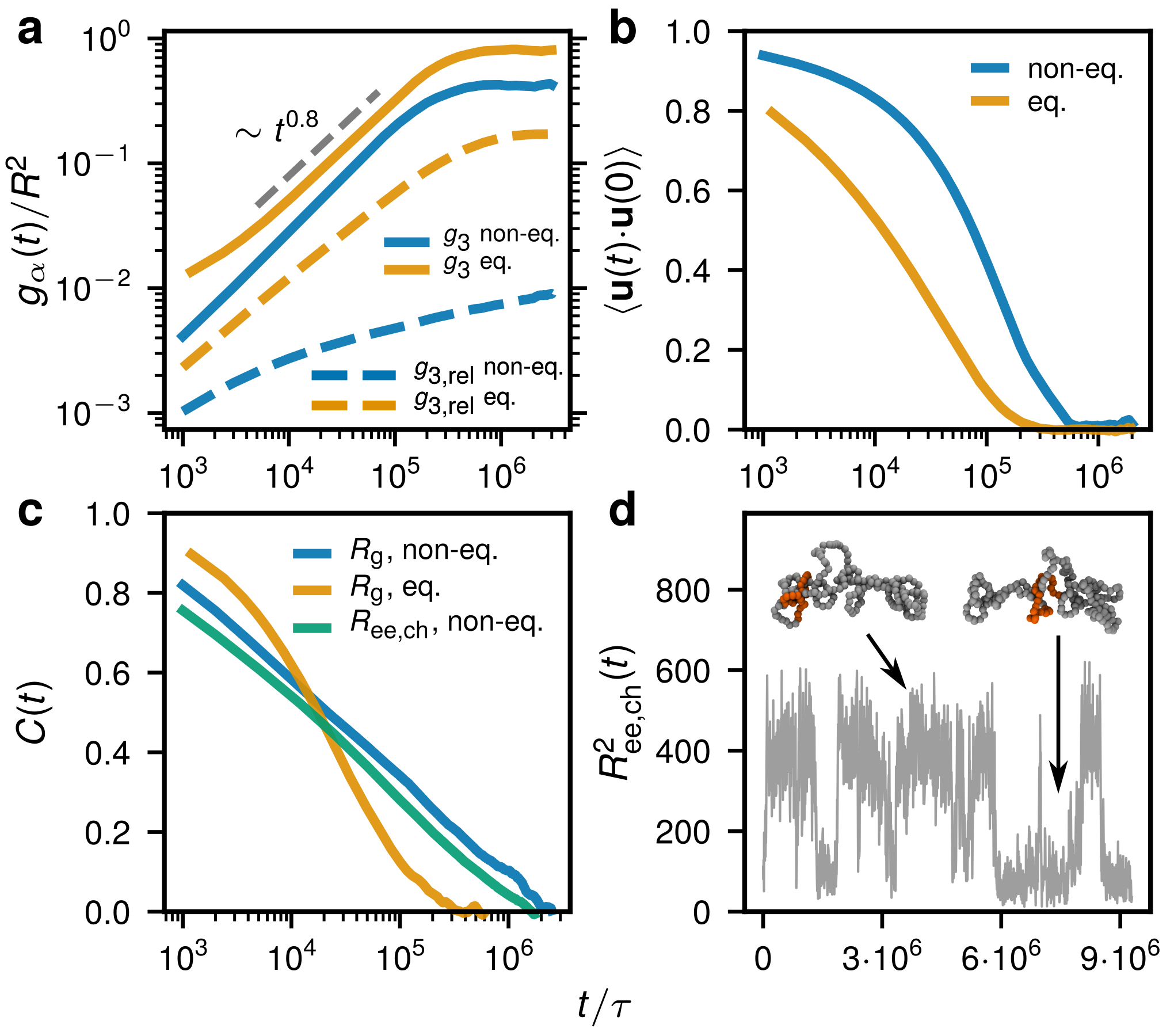}
\caption{\label{fig:dynamics} {\bf Dynamics and relaxation.}
Comparison of the non-equilibrium system (blue) with the equilibrium (yellow).
{\bf a}, Mean-square displacements of the ring's center of mass
$g_{3}$ normalized by the squared sphere's radius $R^2$ of $N = 200$ (solid lines). The relative mean-square displacement $g_{3,\rm{rel}}(t)$ (dashed).
{\bf b}, Terminal autocorrelation function.
{\bf c}, Normalized autocorrelation function for $R_{\rm g}^2$. For comparison the autocorrelation of the squared
``end-to-end" (i.e. between
an active and a passive monomer separated by segment length $N/2$) distance $R^{2}_{\rm ee, ch}$ is shown (green). In all cases we subtract the mean value squared and normalize the autocorrelation functions to unity at time zero.
{\bf d}, Time dependence of $R^2_{\rm ee, ch}(t)$ for one of the rings illustrates tank treading motion.}
\end{figure}

In Fig.~\ref{fig:dynamics} we report dynamical and relaxation
properties of rings in the system with $N = 200$ (averaged over 10 
independent runs). We focus on the late stage dynamics by
discarding the initial period of length $3{\cdot}10^6 \tau$, where major configurational 
rearrangements occur. The late-stage (steady-state) dynamics of the ring's center of 
mass, $g_3(t)$, (Eq.~(S4) in the SI), is much slower than in the equilibrium case \cite{ConfRings2020}, with negligible relative displacements between the rings $g_{3,\rm{rel}}(t)$, (Eq.~(S5)) as shown in Fig.~\ref{fig:dynamics}{\bf a}. Importantly, the latter quantity is invariant under overall constant global rotations and shows that the relative ring's motion essentially stalls. The systems with longer rings display the same behavior.
In confined systems, $g_3(t)$, saturates at a constant value; for the rings with $N = 200$, we find that $g_3(t \rightarrow \infty) \approx 0.4 R^2$, which
is about two times smaller than in the equivalent equilibrium case (Fig.~\ref{fig:dynamics}{\bf a}). This arises mostly from extremely elongated 
and practically frozen rings conformations, due to which the exploration
of the available volume is significantly suppressed (Supporting Video 2). 

We characterize the ring structural relaxation by considering
the terminal autocorrelation function (TACF)
 $\langle \mathbf{u}(t)\cdot\mathbf{u}(0) \rangle$, where $\mathbf{u}(t)$ is the unit vector connecting two monomers
separated by contour distance $N/2$, and the average is taken over all such
monomer configurations within a ring and time \cite{Tsalikis_RL_blends_MAMOL20,Halverson_dynamics2011}. The full decorrelation
time of the TACF ($\approx 6{\cdot}10^5\tau$) 
is about three times longer than in the counterpart 
equilibrium case (Fig.~\ref{fig:dynamics}{\bf b}). In the steady state,
the rings are found in a heavily threaded arrangement with their configurations
being essentially frozen, as evidenced by the static properties.
Since in the steady state the relative ring displacements are
marginal (Fig.~\ref{fig:dynamics}{\bf a}), $\langle \mathbf{u}(t)\cdot\mathbf{u}(0) \rangle$ can decorrelate either through 
internal conformational ring relaxation or collective system rotations.

In what follows, we show that the main pathway that contributes
to the decorrelation of the TACF are correlated, stochastic rotations of the whole system.
The other possible decorrelation mechanism is the internal ring rearrangements, caused by the local explorations of the hot segments or tank treading motion. To show that these do not dominate, we show in Fig.~\ref{fig:dynamics}{\bf c} that 
the normalized autocorrelation function for the ring's $R^2_{\rm g}$, $C(t)$, decorrelates
at a much later time ($\approx 2{\cdot}10^6\tau$) and features a three
decades long logarithmic decay. 
This contrasts with the equilibrium 
behavior, where both structural quantities $R^2_{\rm g}$ 
and $\langle \mathbf{u}(t){\cdot}\mathbf{u}(0) \rangle$ 
decorrelate at 
about the same time (yellow curves in Fig.~\ref{fig:dynamics}{\bf b}
and Fig.~\ref{fig:dynamics}{\bf c}). Although the size of the rings
remains essentially the same during the TACF relaxation, there remains
a possibility of tank treading motion that can significantly impact
the TACF decorrelation but keep the overall size given by $R_{\rm g}$
fixed. As highlighted in Fig.~\ref{fig:dynamics}{\bf d}, the tank treading, a tangential motion of the hot segment along the ring's contour, is indeed observed. Thus, such a mechanism can enhance active-passive phase 
separation when both diffusional and conformational relaxation
of polymers is not possible.
We show that the tank treading does not significantly impact the terminal relaxation by computing the autocorrelation function
for the squared end-to-end distance $R^2_{\rm ee, ch}$ 
between a hot and a cold monomer contour-wise $N/2$ apart (Fig.~\ref{fig:dynamics}{\bf c}). Although it decorrelates slightly faster than the one for $R_{\rm g}^2$, its relaxation time 
is still much larger than that of the TACF. Therefore, collective, stochastic rotations provide the dominant 
contribution to the TACF decorrelation, whereas its relaxation 
time scale can be used as an estimate for rotational diffusion time
(the presence of such global, correlated 
rotations is visible in both Supporting Videos).
Global rotations lead to correlated particle displacements as detailed by computing the spatio-temporal correlation function (Section~S5 in the SI).
Finally, such stochastic rotations can also arise spontaneously 
in confined equilibrium systems coupled to a Langevin thermostat
(see discussion in the SI).

Let us now turn to a discussion of a possible connection of the large correlated motions (rotational diffusion) in ATG and coherent motion of chromatin on micron scale, observed in \cite{Zidovska2013,Shaban_Barth_Bystricky_NAR18}, interpreted also as rotations of the nucleus interior \cite{Rotations_of_chromatin}. In the context of chromatin, many different mechanisms can cause large correlated motions \cite{Ganai_NAR14,Agrawal_2017,Hyeon_Thirumalai_PloS18,Zidovska_Shelley_PNAS18,Nuebler_PNAS18,Liu_correlated_motion_with_lamina_arXiv20,Goldstein_Circulation_of_active_matter_PRL12}.
The works \cite{Hyeon_Thirumalai_PloS18,Zidovska_Shelley_PNAS18,Liu_correlated_motion_with_lamina_arXiv20} focus on the spatio-temporal correlations in the dynamics. While the first one uses the thermal-like model of activity, the latter two investigate the effects of active force dipoles. These are coupled with hydrodynamic interaction in \cite{Zidovska_Shelley_PNAS18}, while in \cite{Liu_correlated_motion_with_lamina_arXiv20}, the active force dipoles act on highly crosslinked chromatin connected to deformable lamina. Note that explicit bonds are used in \cite{Liu_correlated_motion_with_lamina_arXiv20,Ganai_NAR14,Agrawal_2017}, which maintain the compact chromatin state in contrast to fractal globule model, where it arises from the uncrossability of the chains. All the works \cite{Hyeon_Thirumalai_PloS18,Zidovska_Shelley_PNAS18,Liu_correlated_motion_with_lamina_arXiv20} find large-scale correlated motions, but of different origins. In \cite{Hyeon_Thirumalai_PloS18}, the correlated domains coincide with the micro-phase separated domains due to preferential intra-domain interaction, and the activity opposes the coherence, similarly to \cite{Nuebler_PNAS18}. Non-monotonic dependence of the correlation length on time lag has been observed \cite{Zidovska2013,Shaban_Barth_Bystricky_NAR18}; it is not clear if it is a general phenomenon (Fig.~2 in \cite{Shaban_Barth_Bystricky_NAR18}) and in contrast to \cite{Hyeon_Thirumalai_PloS18}, the coherence of the motion even at short time lags is larger for the active systems \cite{Zidovska2013,Shaban_Barth_Bystricky_NAR18}. The correlated motion in \cite{Zidovska_Shelley_PNAS18} comes from the coupling of the hydrodynamic flow due to contractile motors and the nematic ordering of the chromatin fiber (not yet observed), with no discernible effect of local topology (unknottedness) of the conformation. The contractile motor activity in \cite{Liu_correlated_motion_with_lamina_arXiv20} generates the correlated motion as a result of a high number of crosslinks between chromatin fiber, and is even enhanced when more crosslinks are used with a deformable nuclear envelope. Last but not least, apart from the role of activity in correlated motions, other, passive mechanisms are possible \cite{Onuchic_PNAS18,Thirumalai_NatComm18}. The latter work also highlights glassy features of the chromatin dynamics, such as dynamic heterogeneity.
Here we show yet another mechanism giving rise to a correlated motion, namely by activity-induced topological interactions that entangle neighboring domains that subsequently have to move in a correlated fashion. 

Despite the similarity of the coherent motion (and other dynamic features \cite{Active_topoglass_NatComm20,Chuang_directed_motion_CurrBiol06,Nagashima_single_nucleosome_dynamics_JCB19,Transcription_loop_Solovei_biorxiv2020}) of the ATG and chromatin, based on the conformational data ($\nu=0.588$, $\gamma = 1.75$), we conclude that the ATG in the present form is inconsistent with the chromatin large-scale conformational data and the fractal globule model ($\nu=1/3$, $\gamma \simeq 1.1$). The conflict of dynamical and large-scale conformational properties is, however, a persistent issue also in other models that aim at elucidating the physical mechanisms rather than the capturing the conformational details \cite{Zidovska_Shelley_PNAS18}. More work is necessary to conclude if other types of topological glass (dynamic correlations arising from entanglements) can be consistent with fractal globule. One option opens up at smaller length scales (below 1Mbp, which roughly corresponds to 10 beads of our largest system), where the chromatin fiber has nontrivial topology (due to cohesin mediated loops) \cite{Loop_extrusion_Fudenberg_CellRep16} and features less compact statistics \cite{vdEngh_Science92,Schiessel_Traffic18}. Simulations with finer resolution and diverse distribution of the active segments would be necessary to give a satisfactory answer. A notable work in this context, \cite{Hyeon_Thirumalai_PloS18}, use active sites distributed along the polymer according to the epigenetic information of a given chromosome that is modelled as an uncrossable chain with initially fractal-globule large-scale conformational properties. However, the work does not report entanglements or conformational changes of the active segments. Despite some active segments being long (20-80 beads), the relatively lower density, in comparison to ours, and a differential interaction of the active and the inactive chromatin types could suppress or obscure the activity-driven conformational changes we report here. This could be also the reason why the work \cite{Hyeon_Thirumalai_PloS18} does not observe the correlation length to depend on the activity level at short times as reported in experiments \cite{Zidovska2013,Shaban_Barth_Bystricky_NAR18}.  Simulations with more accurate chromatin topology or experiment that would trace the chromatin type in 3D simultaneously with dynamics might elucidate the coherence mechanism.

From the materials research perspective, our work shows that the ATG can be efficiently explored at a significantly reduced computational costs in confinement. We characterized the chain static properties and discovered the tank treading relaxation mechanism that, however, does seem to affect the glass stability, but only the phase-separation properties. More detailed understanding of the topological constraints maintaining the ATG stability should be gained in future to experimentally synthesize ATG and fully characterize this novel dynamical transition.

\section{Associated content}

The Supporting Information is available free of charge at [link to be inserted]. \\

\noindent Additional details on the employed polymer model (Section S1); threading detection (Section S2); conformational properties of the rings (Figures S1 and S2, Tables S1-S3); mean-square displacements (Section S4); spatio-temporal correlations in 
the systems (Figure S3).

\section{Acknowledgments}
We are thankful to R.~Barth and H.-P.~Hsu for fruitful discussions. 
JS acknowledges support from the Austrian Science Fund (FWF) through the Lise-Meitner Fellowship No.~M 2470-N28. This work has been supported by the European Research Council under the European Union's Seventh Framework Programme (FP7/2007-2013)/ERC Grant Agreement No.~340906-MOLPROCOMP. 
IC acknowledges Mobility Fellowship provided by the Vienna Doctoral School in Physics (VDSP). 
The authors would like to acknowledge networking support by the COST Action CA17139. 
We are grateful for a generous computational time at Vienna Scientific Cluster and Max Planck Computing and Data Facility.
 This research was supported in part by the National Science Foundation under Grant No. NSF PHY-1748958 and NIH Grant No. R25GM067110. JS and KK acknowledge the program ``Biological Physics of Chromosomes" KITP UCSB, 2020, for providing a discussion forum this work benefited from.

\section{Notes}
The authors declare no competing financial interest.

\section{Data availability}
The relevant data sets generated and/or analyzed 
in the current study are available from the corresponding authors on reasonable request.


\clearpage

\widetext
\clearpage
\begin{center}
\textbf{\Large Supporting Information}
\vspace*{2mm}
\end{center}
\balancecolsandclearpage

\setcounter{equation}{0}
\setcounter{figure}{0}
\setcounter{table}{0}
\setcounter{section}{0}
\makeatletter
\renewcommand{\theequation}{S\arabic{equation}}
\renewcommand{\thefigure}{S\arabic{figure}}
\renewcommand{\thetable}{S\arabic{table}}
\renewcommand{\thesection}{S\arabic{section}}

\makeatletter
\@fpsep\textheight
\makeatother

\section{Model}
\label{sec:Model}

We use the well-established model \cite{KG_model}, in which the excluded volume interaction between any two monomers is described by a repulsive and shifted Lennard-Jones potential
\begin{equation}
U_{\mathrm{LJ}}(r) = 
\left( 4\varepsilon \left[ \left(\frac{\sigma}{r}\right)^{12} - \left(\frac{\sigma}{r}\right)^{6}\right] +\varepsilon \right) \theta( 2^{1/6}\sigma - r)
\label{eq:LJ}
\end{equation} 
where $\theta(x)$ is the Heaviside step function, 
$\sigma$ is the bead's diameter, and $\epsilon$ sets 
the energy scale.
As in Ref.~\cite{ConfRings2020}, the same potential was used 
for the interaction between monomers and the confining sphere of radius $R$ which was set to achieve an overall monomer density $\rho = 0.85\sigma^{-3}$ (see Tab.~\ref{table:shape_recap}).
The polymer bonds were modeled by a finitely extensible nonlinear elastic (FENE) potential
\begin{equation}
U_{\mathrm{FENE}}(r)  = -\frac{1}{2}r_{\mathrm{max}}^{2} K \log\left[1-\left(\frac{r}{r_{\mathrm{max}}}\right)^{2}\right],
\label{eq:FENE}
\end{equation}
where $K=30.0\varepsilon/\sigma^{2}$ and $r_{\text{max}}=1.5\sigma$. 
These parameters make the chains essentially noncrossable. We also used the angular bending potential
\begin{equation}
U_{\mathrm{angle}} = k_{\theta} (1 - \cos(\theta - \pi))
\label{eq:angular}
\end{equation}
with the parameter $k_{\theta} = 1.5 \varepsilon$ to induce higher stiffness that corresponds to a lower entanglement length $N_{\text{e}} = 28\pm 1$ at the studied monomer density $\rho$ \cite{Halverson_statics2011}. 

Our simulations start from well-equilbrated configurations of 
completely passive ring polymer melts in spherical confinement
produced in Ref.~\cite{ConfRings2020}. Each system contains $M = 46$
ring polymer chains, each of length $N$ ($N = 200, 400, 800$ and $1600$,
corresponding to chain entanglement number $Z = N/N_{\rm e} = 7, 14, 28$ and 57). The choice of $M=46$  chains was inspired by the $23$ pairs of chromosomes in the human diploid cell nucleus, but the main reason is to demonstrate the existence of a topological glass in a small systems to ease future exploration of the phenomenon. At time $t = 0$, the activity was introduced by
coupling a consecutive segment of length $N/8$ on each ring to a Langevin thermostat
at temperature $T_{\rm h} = 3.0 \epsilon$, whereas the rest of the ring
is still maintained at $T_{\rm c} = 1.0\epsilon$ by another Langevin heat
bath. We choose this value of $T_{\rm h}=3T_{\rm c}$, despite the experimental indications of active fluctuations being only about twice the thermal fluctuations. The reason is the heat flux between the active and the passive constituents establishes effective temperatures that are in between the temperatures set by the thermostat. The effective temperatures (measured by the mean kinetic energy) would be the ones measured in the experiments and have the correct ratio about 2 \cite{Topo2}. The equation of motion of the systems were integrated using the 
LAMMPS simlation upackage \cite{LAMMPS} with the time step
$\Delta t = 0.005\tau$ and the damping constant $\gamma = 2/3 \tau^{-1}$,
where $\tau = \sigma (m/\varepsilon)^{1/2}$.

The Langevin thermostat
in spherical confinement can induce stochastic values of 
angular momentum that affect the real dynamics of the system.
This effect can be neutralized by zeroing periodically the total angular momentum 
during the simulations as done in equilibrium simulations in~\cite{ConfRings2020}.
In the present case, unlike in the equilibrium simulations, we do not perform this operation due to
a non-equilibrium character of the studied system as well as 
potential global flows that can arise in active matter states. 
When compared to dynamic equilibrium quantities across this work,
we also used trajectories produced in a similar fashion without 
zeroing the angular momentum. We note, however, that the difference 
in dynamic relaxation times in equilibrium simulations with and without zeroing the angular momentum is rather small.

\begin{figure}
\centering
\includegraphics[width=0.45\linewidth]{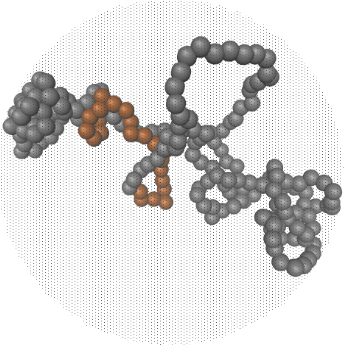}
\includegraphics[width=0.45\linewidth]{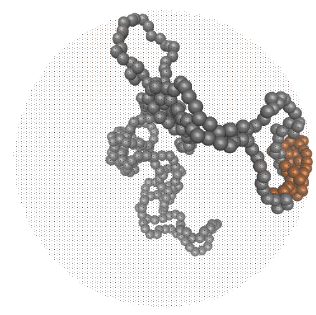}\\
\includegraphics[width=0.45\linewidth]{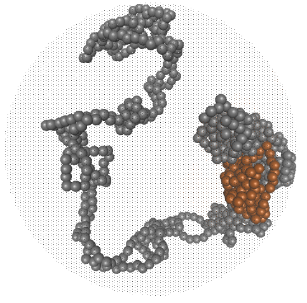}
\includegraphics[width=0.45\linewidth]{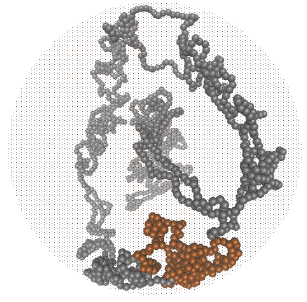}
\caption{\label{fig:snapshots} {\bf Typical conformations of partly
active rings in spherical confinement.} The snapshots correspond
to $N = 200$ (top left), 400 (top right), 800 (bottom left), and
1600 (bottom right).}
\end{figure}

\begin{table*}[ht!b]
		\centering
		\begin{tabular}{||c c | c | c | c | c | c | c | c | c ||} 
			\hline
			$N$  & $N_{\rm h}$  & $R /\sigma$& 
			$\langle R_{\rm g}^{2} \rangle/\sigma^2$&
			$\langle R_{\rm ee}^{2} \rangle/\sigma^2$&
			$\langle R_{\rm ee}^{2} \rangle / \langle R_{\rm g}^{2} \rangle$ &
			$\langle \lambda_{1} \rangle / \langle \lambda_{3} \rangle$ & 
			$\langle \lambda_{2} \rangle / \langle \lambda_{3} \rangle$ & 
			$\langle R_{\rm g}^{2} \rangle / R^2$ &
			$\langle R_{\rm ee}^{2} \rangle / R^2$ \\ \hline\hline
			200 & 25 & 13.72 & 62.4(0.7) & 164.8(6.2) & 2.64 & 12.0(0.7) &
			4.3(0.2) & 0.33 & 0.87 \\ 
			\hline
			400 & 50 & 17.29 & 129.3(0.6) & 304.4(6.2) & 2.34 & 6.5(0.4) &
			3.1(0.7) & 0.41 & 1.01\\
			\hline
			800 & 100 & 21.78 & 227.7(0.5) & 468.6(3.1) & 2.05  & 4.6(0.2) &
			2.7(0.8) & 0.47 & 0.98\\
			\hline
			1600 & 200 & 27.44 & 376.1(0.7) & 810.8(4.7) & 2.15 & 3.5(0.5) & 
			2.2(0.1) & 0.49 & 1.07\\
			\hline
		\end{tabular}
	\caption{{\bf Size and shape properties of partially active rings in a confining sphere}. The mean values as well as their standard errors (indicated
	in the parentheses) were estimated in the steady states.
$R$ is the radius of the sphere. $\langle R_{\rm g}^{2} \rangle$ and 
$\langle R_{\rm ee}^{2} \rangle$ are the mean-square radius of gyration and the mean-square spanning distance between monomers $N/2$ apart, respectively. $\lambda _i$ ($i = 1,2,3$, $\lambda_1 \geq \lambda_2 \geq \lambda_3$) are the eigenvalues of the gyration tensor.}
	\label{table:shape_recap}
\end{table*}
\begin{table*}[h!tb]
\centering
\begin{tabular}{||cc|c|c|c|c|c||}
\hline
$N$  & $N_{\rm{h}}$ &
$\langle R_\text{g}^2 \rangle/\sigma^2$ & 
$\langle R_\text{e}^2 \rangle/\sigma^2$ & 
$\langle R_{\rm ee}^{2} \rangle / \langle R_{\rm g}^{2} \rangle$ &
$\langle \lambda_1 \rangle / \langle \lambda_3 \rangle$ & 
$\langle \lambda_2 \rangle / \langle \lambda_3 \rangle$ \\ \hline\hline
100  & 13 & 18.1(0.1)   & 54.9(0.1) & 3.0 & 7.3(0.1)  & 2.34(0.01) \\\hline
200  & 25 & 65.2(0.3)   & 203.5(3.8) & 3.1 & 12.4(0.1) & 2.81(0.01) \\\hline
400  & 50 & 182.1(0.7)  & 566.1(2.1) & 3.1 & 14.2(0.2) & 3.03(0.02) \\
\hline
\end{tabular}
\caption{{\bf Size and shape properties of the partially active rings in bulk}.
$\langle R_\text{g}^2 \rangle$ is the mean-square radius of gyration,
$\langle R_\text{e}^2 \rangle$ is the mean-square distance between two monomers separated 
by the contour length $N/2$, and $\lambda_i$, $i = 1, 2, 3$ are the eigenvalues 
of the gyration tensor ordered such that 
$\lambda_1 \geq \lambda_2 \geq \lambda_3$.
The value in the parentheses indicates the standard error. For comparison with equilibrium values, please see Table \ref{table:shape_recap_eqconfined}. Adapted from Ref.~\cite{Topo2}.}
\vspace*{2mm}
\label{table:shape_recap_bulk}
\end{table*}

\begin{table*}[htb!]
		\centering
		\begin{tabular}{||c | c | c | c | c | c | c ||} 
\hline
$N$  & $R/\sigma$ &
$\langle R_\text{g}^2 \rangle/\sigma^2$ & 
$\langle R_\text{e}^2 \rangle/\sigma^2$ & 
$\langle R_{\rm ee}^{2} \rangle / \langle R_{\rm g}^{2} \rangle$ &
$\langle \lambda_1 \rangle / \langle \lambda_3 \rangle$ & 
$\langle \lambda_2 \rangle / \langle \lambda_3 \rangle$ \\ \hline\hline
200  & 13.72 & 26.4(0.2)   & 73.4(0.6) & 2.8  & 5.64(0.04) & 2.25(0.01) \\\hline
400  & 17.29 & 44.4(0.7)   & 120.7(2.5) & 2.7 & 5.24(0.08) & 2.14(0.02) \\\hline
800  & 21.78 & 73.1(1.1)   & 195.4(3.8) & 2.7 & 4.93(0.10) & 2.06(0.01) \\\hline
1600 & 27.44 & 120.5(2.8) & 320.2(10.4) & 2.7 &  4.89(0.12) & 2.03(0.02)  \\
\hline
\end{tabular}
\caption{{\bf Size and shape properties of the equilibrium confined rings}.
$R$ is the radius of the confining sphere,
$\langle R_\text{g}^2 \rangle$ is the mean-square radius of gyration,
$\langle R_\text{e}^2 \rangle$ is the mean-square distance between two monomers separated 
by the contour length $N/2$, and $\lambda_i$, $i = 1, 2, 3$ are the eigenvalues 
of the gyration tensor ordered such that 
$\lambda_1 \geq \lambda_2 \geq \lambda_3$.
The value in the parentheses indicates the standard error. Adapted from
Ref.~\cite{ConfRings2020}}
\label{table:shape_recap_eqconfined}
\end{table*}

\section{Threading detection}
We detect the threadings using a minimal surface method that has been used successfully to analyze threading constraints for systems containing ring polymers in equilibrium  \cite{Smrek_Grosberg_minimal_surfaces_ACS16,ConfRings2020,Tadpoles2020} and out of equilibrium \cite{Smrek_Kremer_Rosa_ACS19,Active_topoglass_NatComm20,Topo2}. The essence of the method is an unambiguous definition of the threading as the intersection of a rings contour with a disk-like surface spanned on another ring. As there are many possible surfaces with the contour of the ring a surface of minimal surface area is chosen. For the details on practical implementation of the algorithm we refer the reader to  \cite{Smrek_Kremer_Rosa_ACS19,Active_topoglass_NatComm20}. The reported value of number of threaded neighbors $n_{\rm{tn}}$ is a mean number of rings each ring threads as in \cite{Active_topoglass_NatComm20}. 

\section{Additional conformational properties}
\begin{figure*}
\centering
\vspace*{3cm}
\includegraphics[width=0.49\linewidth]{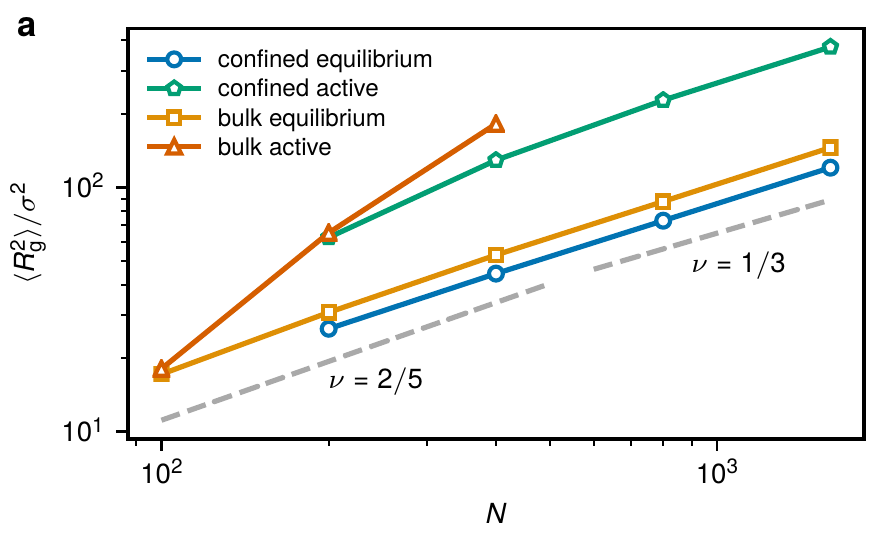}
\includegraphics[width=0.49\linewidth]{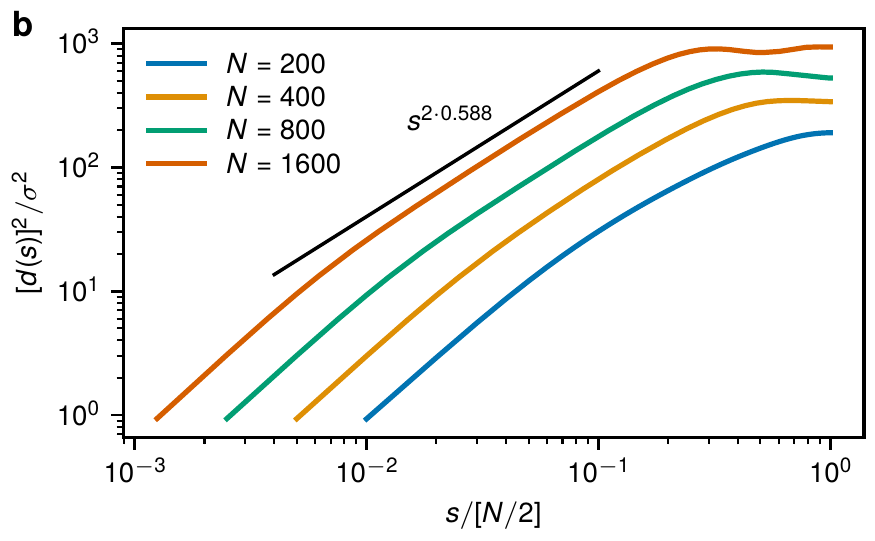}
\includegraphics[width=0.49\linewidth]{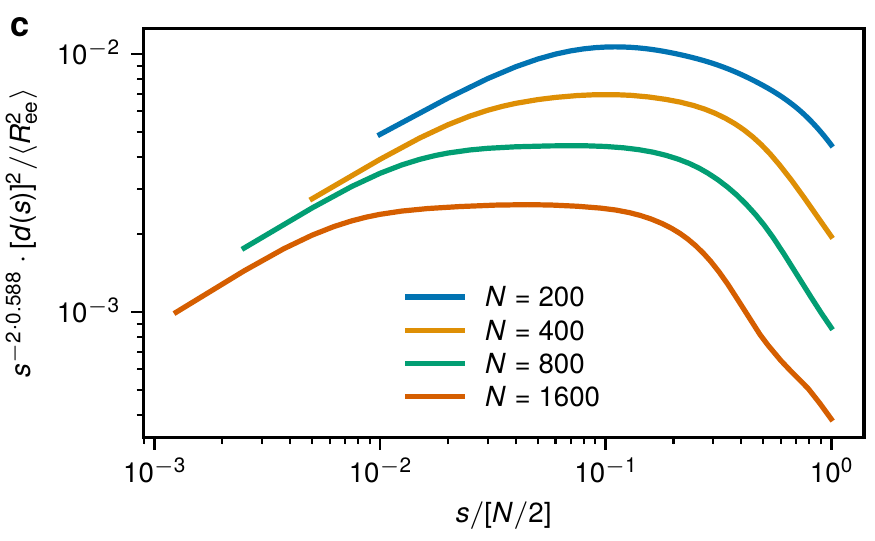}
\includegraphics[width=0.49\linewidth]{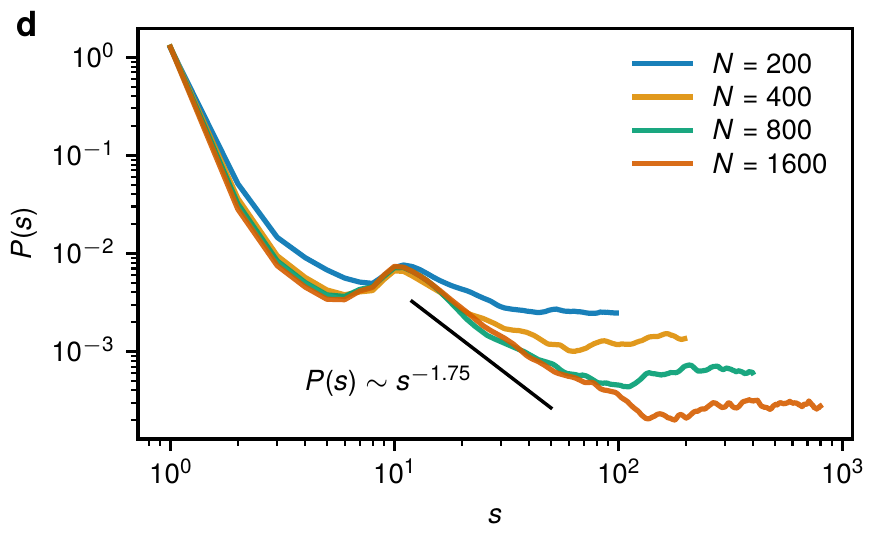}
\caption{\label{fig:additional_shape_props} {\bf Additional conformational properties.} {\bf a}, Comparison of the scaling of the radius of gyration with the ring length for different systems. The confined active rings are from the present work, the confined equilibrium rings are from \cite{ConfRings2020}, the bulk equilibrium are from \cite{Halverson_statics2011}, and the bulk active rings from \cite{Active_topoglass_NatComm20}. The equilibrium scaling exponent $\nu=1/3$ is shown as well as the crossover with ``effective exponent" $2/5$. The seeming compact scaling of the active confined rings is not due to their internal structure (see {\bf b}), but just because the confining radius $R$ scales with $N^{1/2}$ as the systems of different $N$ were simulated with the same number of chains and the same density. {\bf b}, The mean-square internal-distance $d(s)$ is computed as the mean square end-to-end vector of a segment of length $s$ averaged over its position within a ring and over different rings in the steady state. In the intermediate distances ($s/(N/2) \in [10^{-2};10^{-1}]$) we recover self-avoiding walk scaling exponent $0.588$ for the longer ($N\geq 400$) rings. These exhibit also monotonic profile for large contour distances. {\bf c}, The mean-square internal-distance $d(s)$ rescaled by the $s^{0.588}$. The broadening plateau for the rings of $N\geq400$ shows the asymptotic self-avoiding regime. {\bf d}, The contact probability $P(s)$ is the probability of finding the endpoints of a segment $s$ at distance below $2^{1/6}\sigma$. It is an average over the segment's position within a ring and over different rings in the steady state. At intermediate distances and for long rings we recover exponent $\gamma$ close to $1.75$ consistent with the self-avoiding random walk configuration. At longer lengths, $P(s)$ goes to a constant signifying the positional decorrelation due to reflections from the wall --- this is typical profile of an equilibrium globule, i.e. confined melt of linear chains. The exponent $\gamma$ is smaller for shorter rings. The non-monotonic character is due to the phase separation of the hot and cold segments and the doubly-folded structure.}
\end{figure*}

In Fig.~\ref{fig:additional_shape_props} we report additional conformational properties of the rings. There we compare scaling of the gyration radius in the present active confined system with the bulk and the equilibrium counterparts. Additionally we show the profile of the contact probability and the scaling of the mean square internal distance that exhibits the self-avoiding statistics.

\section{Mean-square displacements}
We compute the mean-squared displacement $g_{3}(t)$ as
\begin{equation}
\label{eq:g3_time_avg}
g_3(t,t_0,t_{\rm tot}) = \left\langle \frac{1}{t_{\rm tot}-t} \int_{t_0}^{t_0+t_{\rm tot}-t} 
\left[ \mathbf{R}(t' + t) - \mathbf{R}(t') \right]^2 \textrm{d} t' \right\rangle
\end{equation}
where $t_{0}$ is the initial time point chosen as the onset of the steady state ($3{\cdot}10^{6}\tau$ in case of active rings and $0$ for equilibrium), $t_{\rm tot}$ is the total simulation time, $\mathbf{R}$ is the position of the ring's center of mass with respect to the global center of mass and the angles mean averaging over different rings. 

Additionally, we compute the relative mean-square distance $g_{3,\rm{rel}}(t)$:
\begin{equation}
\label{eq:g3rel_time_avg}
g_{3,\rm{rel}}(t) = \left\langle\frac{1}{t_{\rm tot}-t} \int_{t_0}^{t_0+t_{\rm tot}-t} 
\left[ d_{ij}(t' + t) - d_{ij}(t') \right]^2 \textrm{d} t' \right\rangle_{ij}
\end{equation}
where $t_{0}$ and $t_{\rm tot}$ are as above, $d_{ij}$ is the relative \emph{distance} between rings $i$ and $j$ and the $\langle\ldots\rangle_{ij}$ is the average over all possible ring pairs in the system.

\section{Spatio-temporal correlations}
\label{sec:spatio-temporal_correlation}
\begin{figure*}
\vspace*{5cm}
\centering
\includegraphics[width=0.49\linewidth]{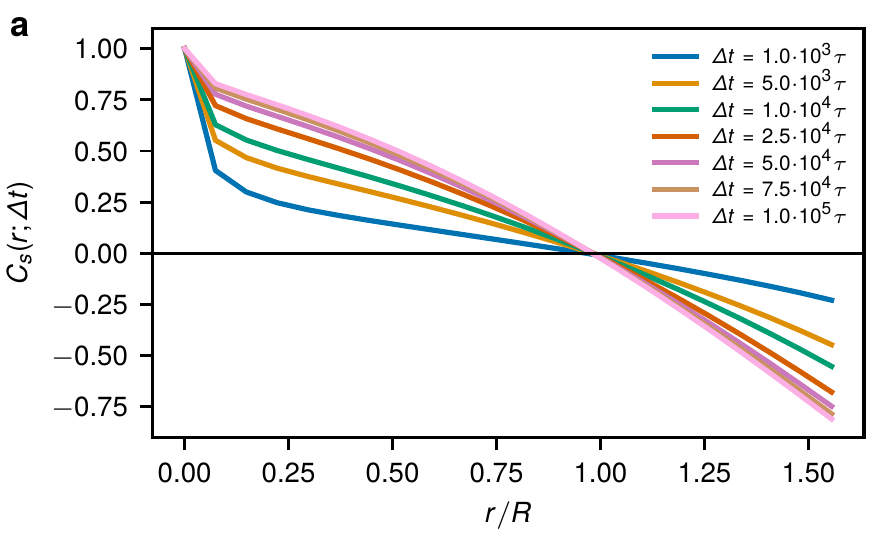}
\includegraphics[width=0.49\linewidth]{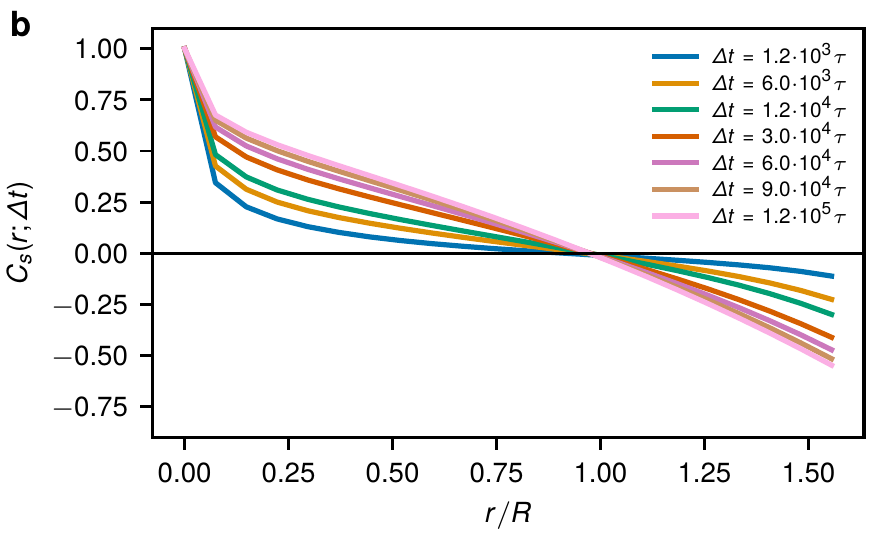}
\includegraphics[width=0.49\linewidth]{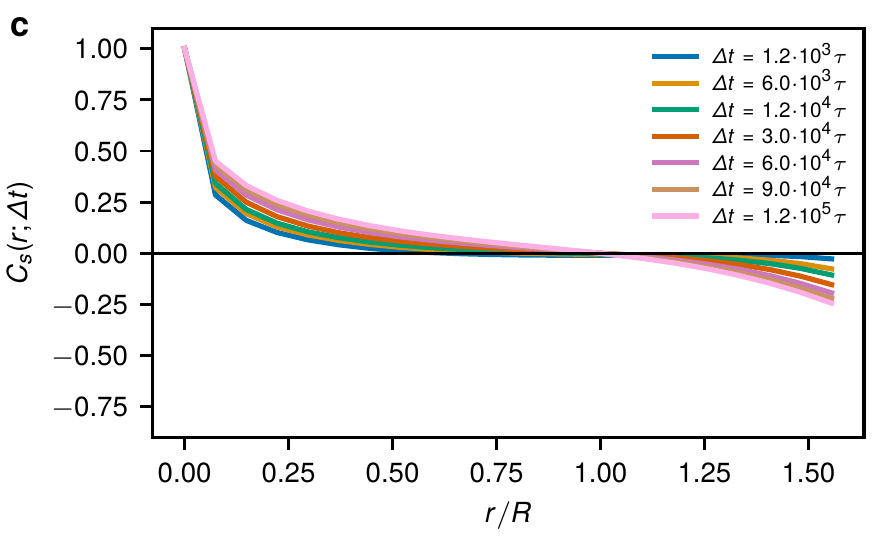}
\caption{\label{fig:dcs_correlations} {\bf Spatio-temporal displacement correlation.} The correlation for $N=200$ system computed by \eqref{eq:Crt} for {\bf a} the active confined rings, {\bf b} equilibrium rings without zeroing the angular momentum, {\bf c} equilibrium rings with zeroing the angular momentum from Ref.~\cite{ConfRings2020}.}
\end{figure*}

The spatio-temporal correlations are computed similarly to \cite{Zidovska2013,Hyeon_Thirumalai_PloS18} as 
\begin{equation}
\label{eq:Crt}
C_{\rm s}(r; \Delta t) = \left\langle \frac{\sum_{i>j}[\Delta\mathbf{R}_{i}(t,\Delta t)\cdot\Delta\mathbf{R}_{j}(t,\Delta t)]\delta(\mathbf{R}_{ij} - r)}{\sum_{i>j}\delta(\mathbf{R}_{ij} - r)} \right\rangle
\end{equation}
where $\Delta\textbf{R}_{i}(t,\Delta t)$ is the displacement of the $i$-th monomer in lag time $\Delta t$ as measured in time $t$. The angular brackets represent averaging over time, in the active case only over the steady state. In the active system, the correlation decays significantly slower in comparison to equilibrium and there is a strong anticorrelation at longer lag times at the opposing positions in the spherical confinement ($r\simeq 1.5R$) (Fig.~\ref{fig:dcs_correlations}{\bf a}). In part, this is a consequence of the Langevin dynamics that induces stochastic angular momentum also in equilibrium (Fig.~\ref{fig:dcs_correlations}{\bf b}). However, the anticorrelation is much more pronounced in the active topological glass state, and almost nonexistent in equilibrium with zeroed angular momentum (Fig.~\ref{fig:dcs_correlations}{\bf c}). We observe larger correlation length for the active case consistent with \cite{Zidovska2013,Shaban_Barth_Bystricky_NAR18,Liu_correlated_motion_with_lamina_arXiv20}, but in contrast to \cite{Hyeon_Thirumalai_PloS18}. However, the correlation length seems to be monotonically increasing and saturating with time that is consistent with some cases in \cite{Shaban_Barth_Bystricky_NAR18}, but non-monotonic correlation length has been observed in other cases at longer time lags \cite{Shaban_Barth_Bystricky_NAR18,Zidovska2013,Liu_correlated_motion_with_lamina_arXiv20,Hyeon_Thirumalai_PloS18}.

\balancecolsandclearpage

\balancecolsandclearpage

\balancecolsandclearpage

\end{document}